\documentclass[showpacs,prb,twocolumn,superscriptaddress]{revtex4}
\usepackage{graphicx}
\newcommand{\etal}{{\it et al.}}

\begin{document}

\title{X-ray natural dichroism and chiral order in underdoped cuprates}

\author{M. R. Norman}
\affiliation{Materials Science Division, Argonne National Laboratory, Argonne, IL 60439, USA}

\begin{abstract}
The origin of the Kerr rotation observed in the pseudogap phase of cuprates has been the subject
of much speculation.  Recently, it has been proposed that this rotation might be due to chiral charge
order.  Here, I investigate whether such order can be observed by x-ray natural circular
dichroism (XNCD).  Several types of charge order were considered, and they can give
rise to an XNCD signal depending on the stacking of the order along the c-axis.
\end{abstract}

\date{\today}

\maketitle
Since its observation over twenty years ago, the nature of the pseudogap phase in underdoped
cuprates has been a subject of much debate \cite{timusk,NPK}.  Time reversal symmetry
breaking was proposed by Varma \cite{varma}, and based on this prediction,
angle resolved photoemission (ARPES) experiments were done by Kaminski \etal~\cite{kaminski} 
on underdoped Bi$_2$Sr$_2$CaCu$_2$O$_{8+\delta}$ (Bi2212) that
did indeed observe a dichroism signal by using circularly polarized light.  Their signal was consistent
with an order parameter like temperature evolution that set in below the pseudogap
temperature T$^*$.  These results,
though, have yet to be confirmed by other groups \cite{borisenko}.  Moreover, even if
confirmed, such a signal could instead be due to chiral order or certain types of structural
distortions rather than time reversal breaking.  On the other hand, subsequent elastic neutron scattering
experiments did observe time reversal breaking in underdoped YBa$_2$Cu$_3$O$_{6+x}$
(YBCO) \cite{bourges}
that sets in below T$^*$.  This has now been seen in a variety of cuprates \cite{bourges-rev}.
Interestingly, this symmetry breaking has not been seen by $\mu$SR \cite{luke}.

After the neutron results, a novel Sagnac interferometry experiment was done by Kapitulnik's
group \cite{xia} that observed a Kerr rotation that sets in at a temperature below that where the
elastic magnetic signal is first seen by neutrons.  In Bi$_2$Sr$_2$CuO$_{6+\delta}$ (Bi2201),
a similar signal was seen that set in at the same temperature ARPES measurements saw
an energy gap develop \cite{he}.  Originally, it was suggested that this signal was a signature
of a tiny ferromagnetic moment, but as Orenstein has pointed out \cite{joe}, a Kerr rotation
was seen in the antiferromagnet Cr$_2$O$_3$ where a structural distortion accompanies
the magnetic order \cite{krich}.  This occurs since although $<P>$ and $<M>$ vanish
because of the staggered order (where $P$ is the polarization and $M$ the magnetization),
$<(P \cdot M)_{\perp}>$ is non-zero where $\perp$ denotes the components of the scalar
product perpendicular
to the propagation vector of the light.  That is, the structural distortion flips sign with the moment,
and so the scalar product is invariant to the staggering.
A different magneto-chiral phase has recently been advocated by Varma and collaborators 
based on orbital currents to explain the Kerr effect in the cuprates \cite{aji}.
But the idea of chiral charge order has also been proposed \cite{hosur,joe2}.
One motivation for the latter proposal is that the onset of the Kerr rotation in YBCO does not occur
at T$^*$, but rather at a lower temperature where charge order has been seen to develop by
x-ray scattering \cite{keimer,hayden}.

To further explore this, it would be useful to identify other probes that could detect
such chiral order.  XNCD is a natural one to propose \cite{alagna}.  This is the difference in
absorption of left and right circularly polarized light from time even processes.  It is sensitive to inversion
symmetry breaking, and as it is site specific, it can be used to gain information on the spatial
nature of the order.  Interestingly, an XNCD signal was seen at the Cu K edge in Bi2212 \cite{kubota} that
has an order parameter like evolution with temperature that matched the ARPES dichroism
signal.  Though many space group refinements of Bi2212 break inversion symmetry, because of the
presence of glide planes, the XNCD signal due to the crystal structure is zero when the light is directed along the
c-axis \cite{matteo} as in the experiment.  A lower symmetry than simply
the crystal structure would be required to generate an
XNCD signal, and chiral order could indeed satisfy this condition.

In the present paper, I investigate whether such chiral order could be the source of an
XNCD signal, and find that if the order is such as to violate the two glide planes present in
Bi2212 (one perpendicular to the CuO$_2$ planes, the other parallel), then an XNCD signal
does occur.

I start with a discussion of the Bi2212 space group.  The basic unit cell is face centered orthorhombic.
For purposes here, the superstructure modulation observed in Bi2212 is ignored, since
this involves a translation operation that is not relevant to the symmetry considerations
allowing for XNCD.  In our previous study \cite{matteo}, several space group refinements were considered,
but for the purposes of this paper, two will be studied.  The first one (Bb2b) of Gladyshevskii and 
Fl\"ukiger \cite{glady} is non-centrosymmetric, the second one (Bbmb) of Miles \etal~\cite{miles}
is centrosymmetric.  All refinements have in common the existence of two glide
planes (with the glides along the supermodulation b-axis), one perpendicular to the a-axis which runs 
through the planar oxygen ions \cite{foot1}, the other perpendicular to the c-axis (midway between successive
CuO$_2$ bilayer units, which is also midway between the two BiO planes).  One can easily see that any 
given order must violate both glide planes in
order to give a non-zero signal, since XNCD involves a sum over all the atoms in the crystal associated
with the x-ray edge studied.

XNCD originates from interference between dipole (E1) and quadrupole (E2) contributions to x-ray
absorption \cite{alagna}.  This requires inversion symmetry to be broken.  This contribution is a
time even pseudo-deviator \cite{goulon-jetp}, which for most (but not all) symmetries would give rise to optical dichroism as well
(a pseudo-scalar) \cite{chemla}.  There can be a contribution to XNCD
as well from interference between electric (E1) and magnetic (M1) dipoles, but this contribution is
very small, which was verified in the present case by numerical simulations.

Numerical results are generated using the FDMNES program \cite{fdmnes,so}.
The simulations were done using local density atomic potentials (Hedin-Lunqvist
exchange-correlation function) in a muffin tin approximation that considers multiple scattering
of the photoelectron around the absorbing site \cite{natoli,foot2}.
The cluster radius is limited by the photoelectron escape depth \cite{foot3}.
Here, fixed cluster radii of 3.1\AA~and 4.9\AA~are considered as in Ref.~\onlinecite{matteo},
but in one case, larger clusters of radii 5.5\AA~and 6.5\AA~were studied.
For all calculations shown, the incident light is along the c-axis.

\begin{figure}
\includegraphics[width=0.75\hsize]{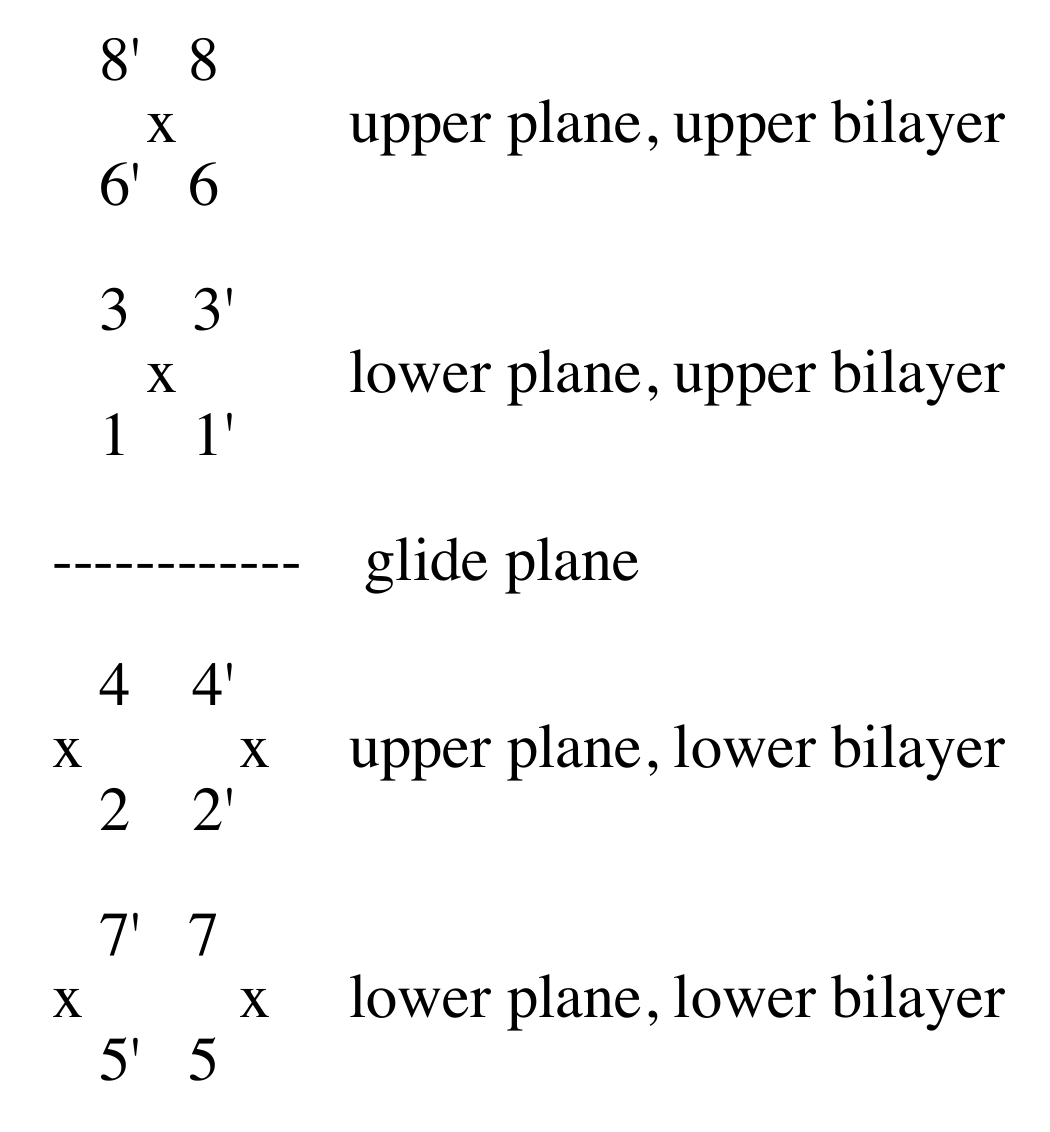}
\caption{Oxygen sites in the four CuO$_2$ planes of the unit cell of Bi2212 (shown are four oxygens per plane).
The numbers refer to the 
space group operations in the text, with the two different planar oxygen sites shown as unprimed and primed.
x denotes the location of the coppers.  The a-axis runs along 1-1$^{\prime}$, the b-axis along 1-3, with a glide plane along
1-3 as well.  The other glide plane ($\perp$c) is shown as the dashed line.} 
\label{fig1}
\end{figure}

To proceed, discussing symmetries of the CuO$_2$ planes for Bi2212 is in order (Fig.~1).
There are two different planar oxygen sites, each of which forms rows along the b-axis, with the 
different rows alternating as one moves along the a-axis.
Associated with this is a glide plane orthogonal to the a-axis.
There are also four CuO$_2$ planes per unit cell, composing an upper bilayer and a lower bilayer,
with another glide plane in between the two bilayers.  The two oxygen rows are interchanged as
one moves from each layer to the next, with the upper plane of the upper bilayer and the lower plane
of the lower bilayer having the same pattern, and the other two planes staggered relative to this by
half a lattice constant along the a-axis.  Note
that as a consequence of these two different planar oxygen sites, even in centrosymmetric
refinements of Bi2212, inversion symmetry is locally broken about the copper sites.
As for the coppers, they stagger differently, translating along the a-axis by half a lattice constant as one goes from
one bilayer to the next.  This can be understood from the eight space group operations,
which are (1) the identity, (2) two-fold rotation (-x,y,-z), (3) first glide (-x,y+1/2,z), (4) second glide (x,y+1/2,-z), 
and then (5-8) the above operations followed by the face centering translation (1/2,0,1/2).  These operations
are used to label the oxygen ions in Fig.~1.

\begin{figure}
\includegraphics[width=0.8\hsize]{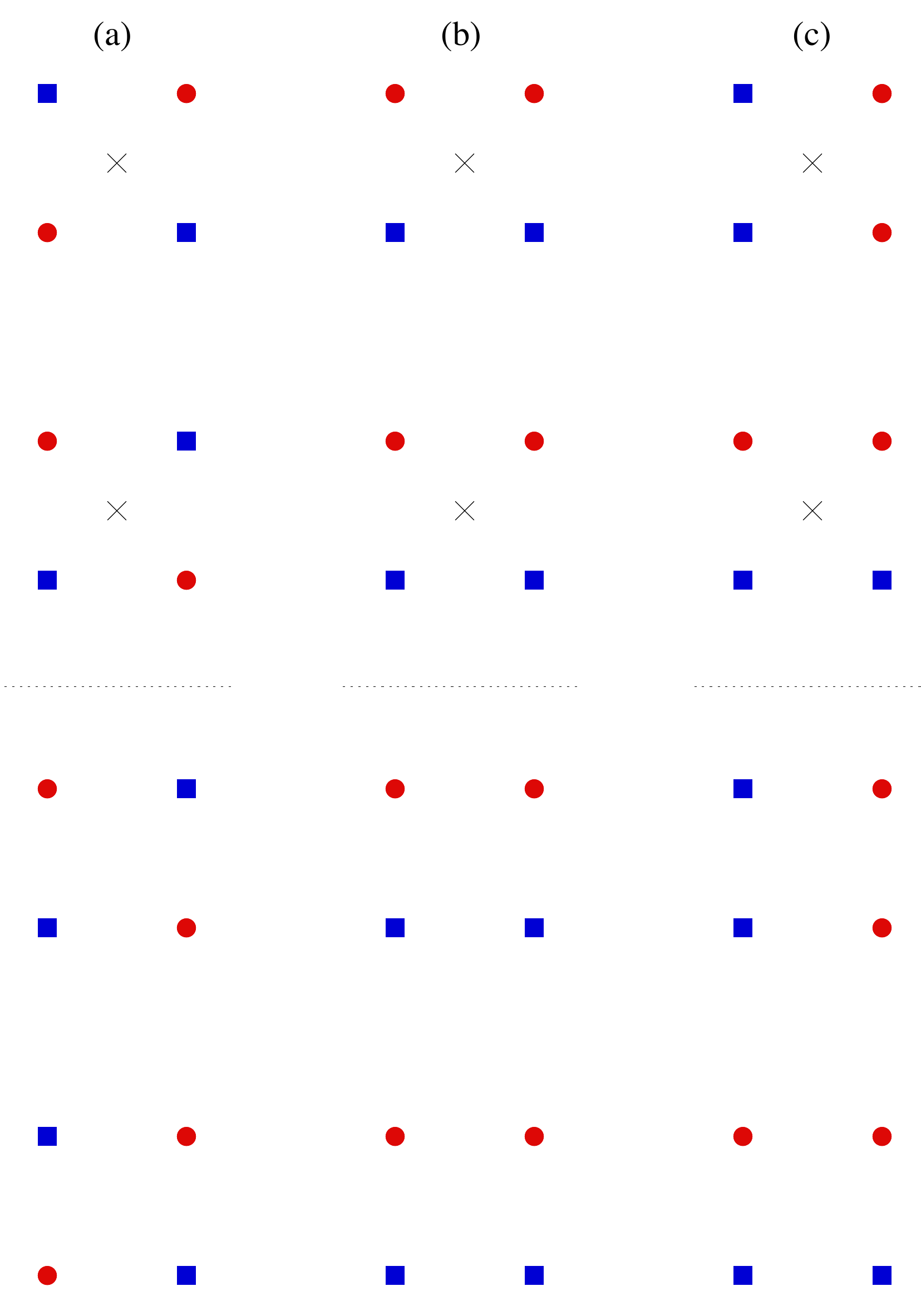}
\caption{(Color online) Three charge patterns considered in the text - (a) pattern 1, (b) pattern 2, and (c) pattern 3.
Circles and squares are oxygen ions, crosses are coppers.  Positions of the ions are as in Fig.~1.
In the calculations, squares have a charge excess of 0.1, circles a charge deficiency of 0.1.}
\label{fig2}
\end{figure}

I will now consider charge ordering patterns on the oxygens that do not break translational symmetry, since this is what the neutron
data indicate \cite{bourges}, and as mentioned above, the translational part of any order is not relevant for the
symmetry considerations allowing for an XNCD signal.  Fig.~2a shows a pattern that is the charge analogue
of the oxygen spin pattern that was considered in Ref.~\onlinecite{bourges} as an alternate to orbital
currents, referred to here as pattern 1.  Note that this pattern breaks the first glide.  Another pattern,
referred to as pattern 2, flips the charges in the second row of oxygen ions (Fig.~2b).  This pattern also violates the
first glide.  To violate the second glide, these two patterns must have the correct staggering along the
c-axis.  For pattern 1, the staggering is (-,+,+,-) where the order refers to the plane sequence
(upper plane, upper bilayer;  lower plane, upper bilayer; upper plane, lower bilayer;  lower plane, lower bilayer)
as in Fig.~1, with the sign referring to that of the oxygen site in the lower left corner of each plane
(6$^{\prime}$,1,2,5$^{\prime}$).  All other stackings with an even number of + and - signs give a vanishing signal.
For pattern 2, a non-zero signal under the same condition of an even number of + and - signs
requires that all planes are in phase (+,+,+,+).
In both cases, the space group reduces to B121 with four operations: the identity, two-fold rotation (-x,y,-z), and
the face centering translation (1/2,0,1/2) of these two.

\begin{figure}
\includegraphics[width=\hsize]{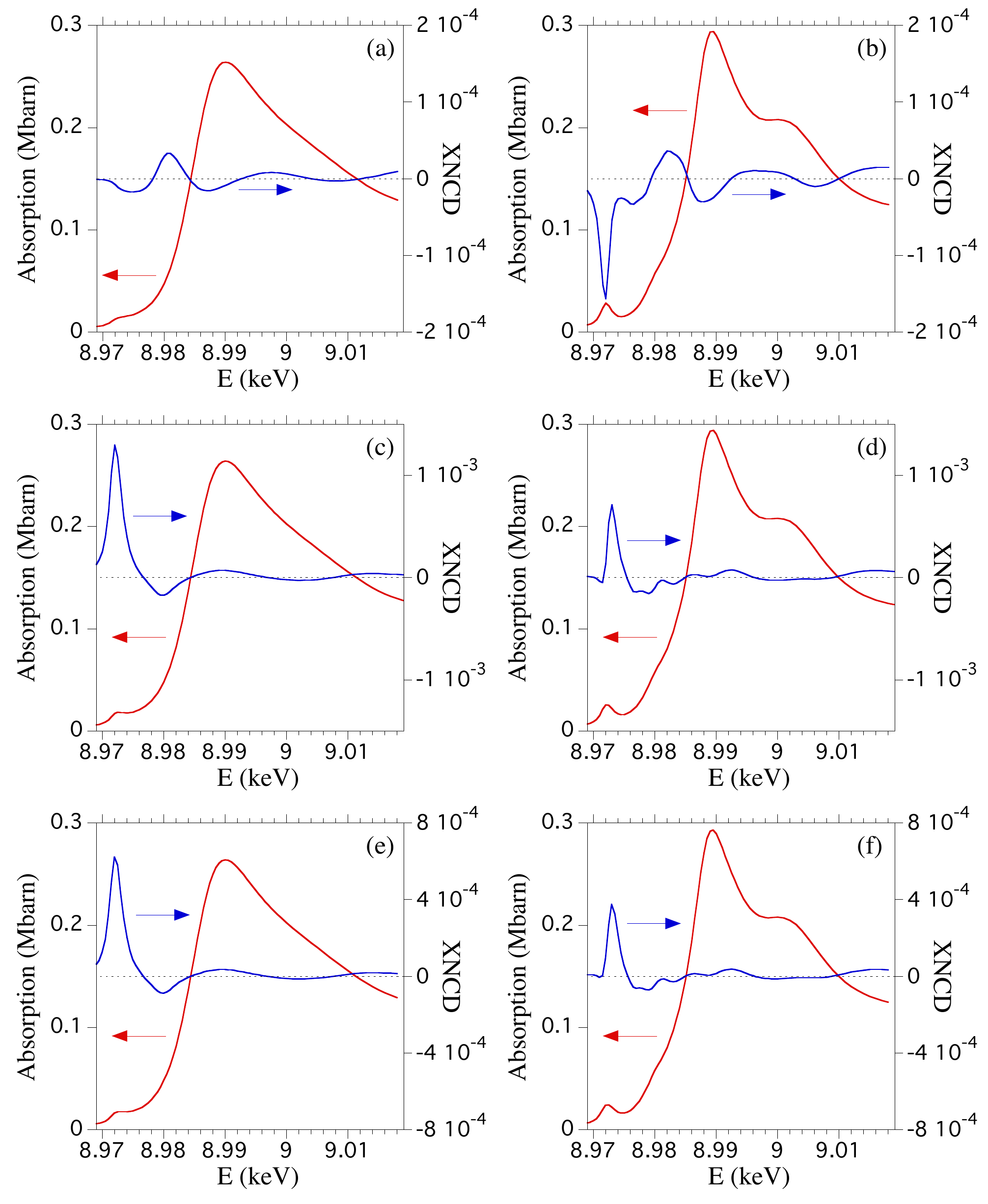}
\caption{(Color online) XNCD for the various charge patterns described in the text at the Cu K edge in Bi2212
(with $k$ along the c-axis) for the crystal refinement of Ref.~\onlinecite{glady}.
Also shown is the absorption signal itself.  The left column is for a cluster radius of 3.1\AA, the right one for 4.9\AA.
The top row is for pattern 1, the middle row for pattern 2, and the bottom row for pattern 3.
Note the differing scales on the right y-axis for each row.}
\label{fig3}
\end{figure}

\begin{figure}
\includegraphics[width=\hsize]{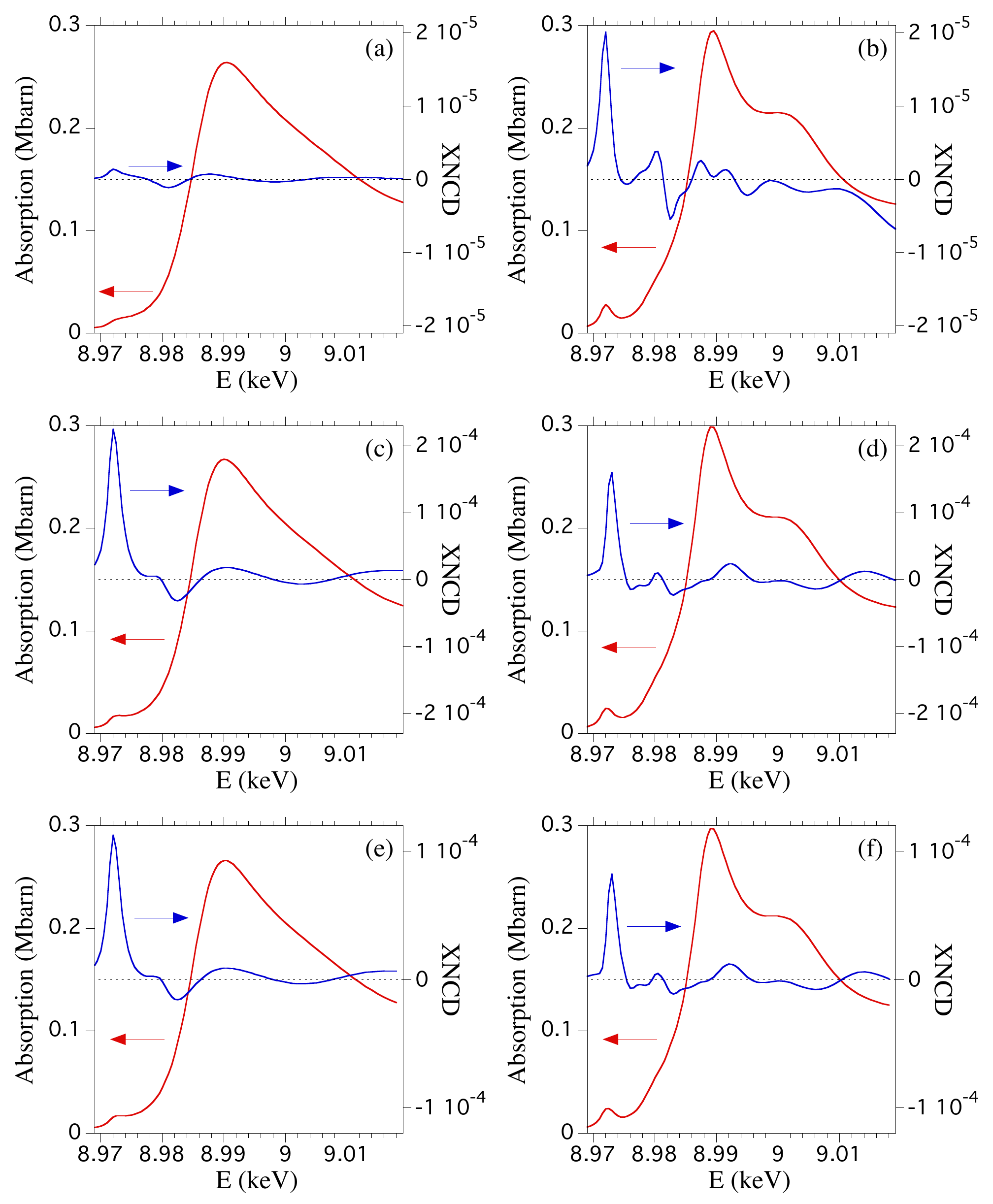}
\caption{(Color online) XNCD for the various charge patterns described in the text at the Cu K edge in Bi2212
(with $k$ along the c-axis) for the crystal refinement of Ref.~\onlinecite{miles}.
Same notation as in Fig.~3.}
\label{fig4}
\end{figure}

Next, a true chiral pattern is considered based on the suggestion of Ref.~\onlinecite{hosur}.
This pattern is the same as in Fig.~2b, except it is rotated around the center of the four oxygen sites
in each plane as one goes from plane to plane, with
a rotation sequence of (-90$^{\circ}$,0$^{\circ}$,90$^{\circ}$,180$^{\circ}$).  This pattern, though, preserves
the symmetry operation (-x+1/2,y+1/2,z+1/2) and thus has a vanishing XNCD signal.  An alternate pattern
presented in Ref.~\onlinecite{joe2} has the rotation sequence (-90$^{\circ}$,0$^{\circ}$,-90$^{\circ}$,0$^{\circ}$).
This pattern (Fig.~2c), denoted as pattern 3,  breaks all symmetries (P1 space group), allowing for XNCD.

Fig.~3 shows the calculated XNCD signal at the Cu K edge for the three patterns with a cluster radius of 
3.1\AA~(a copper and its five surrounding oxygen ions) in the left column, and a cluster radius of 
4.9\AA~(which contains 37 atoms) in the right column.
Spin-orbit was included, but this only had a modest effect.
A charge imbalance of $\pm$0.1 $e^-$ was assumed on the oxygen sites.
Note that flipping the charge pattern flips the dichroism signal.
Complex energy profiles are found.  They differ between the various
charge patterns, and are sensitive to the cluster radius.
Note that the dichroism in pattern 3 is half that of pattern 2.  This can be understood from the symmetry of
the two patterns shown in Fig.~2.
The magnitude of the dichroism is significant, with values up to 4 x 10$^{-3}$ of the absorption maximum, 
which is not only larger than that
claimed in Ref.~\onlinecite{kubota}, but well within the range of detection of modern x-ray sources.
The largest signal typically is seen at the pre-edge peak
at 8.972 keV, expected since this corresponds to $1s - 3d$ excitations (as opposed to the edge itself, which
corresponds to $1s - 4p$ excitations).  This is in contrast to the experimental result, which is simply a
positive peak at 8.99 keV followed by a negative peak at 8.998 keV.  As discussed in
our earlier work \cite{matteo}, the observed signal seems most consistent with a small energy shift between left
and right circularly polarized light, rather than intrinsic dichroism.  Regardless, the XNCD energy profile
calculated here is very sensitive to the nature of the charge order, thus providing a unique fingerprint for this order.

In Fig.~4, results are shown for the centrosymmetric refinement of Miles \etal~\cite{miles}.  Generally, the
dichroism is about an order of magnitude smaller, not surprising since this refinement preserves inversion symmetry.
Larger cluster
radii were also run for pattern 2 (5.5\AA~and 6.5\AA).  The 5.5\AA~result was similar to the 4.9\AA~one, whereas
the 6.5\AA~one had a pre-edge XNCD signal which was partially suppressed.
Still, the XNCD signal is generally large enough that it should be detectable if present.

In summary, I find that XNCD should be an exquisite probe of chiral order in cuprates, assuming that the order is such
to allow a non-zero XNCD signal.  Each charge pattern calculated has a unique energy profile that should be
detectable at modern x-ray sources.  In fact, a variety of potential x-ray measurements could be done that would allow
for a thorough interrogation of such order, including its potential magneto-chiral nature \cite{goulon-jetp,sergio},
and it would be quite interesting if such experiments were attempted on a variety of cuprates.
 
Work at Argonne is supported by Basic Energy Sciences, Office of Science, U.S.~Dept.~of Energy,
under Contract No.~DE-AC02-06CH11357.  The author would like to thank Sergio Di Matteo
for many helpful discussions.


\begin{thebibliography}{99}
\bibitem{foot1}
For non-centrosymmetric refinements, the oxygen ions are slightly displaced off the glide plane.
\bibitem{foot2}
When constructing the potentials, it is assumed that the coppers are in a d$^9$s$^2$ configuration and the oxygens 
in a p$^{4\pm0.1}$s$^2$ configuration.
\bibitem{foot3}
The results presented involve a convolution of the calculated spectrum with 
both a core hole (1.89 eV) and a photoelectron
lifetime, with the latter having a strong energy dependence (at high energies, 15 eV, with a
midpoint value at 30 eV above the Fermi energy).
A Fermi energy (i.e., a core level shift) of -7.4 eV was assumed.
Calculations were also done with a Fermi energy determined by the code, which led to only
minor differences in the pre-edge region.
\end{thebibliography}
\end{document}